\documentclass{article} \usepackage{slashed,epsfig,amsmath,amssymb,enumitem} \usepackage[papersize={8.5in,11in}]{geometry}
   \usepackage{bm}
\usepackage{graphicx}
\newcommand{\be}{\begin{equation}}
\newcommand{\ee}{\end{equation}}
\title{Galaxy Formation in the Early Universe}
\author{J. W. Moffat\\
Perimeter Institute for Theoretical Physics, Waterloo, Ontario N2L 2Y5, Canada\\
and\\
Department of Physics and Astronomy, University of Waterloo, Waterloo,\\
Ontario N2L 3G1, Canada}
\begin{document}
\maketitle


\begin{abstract}Recent observations by the James Webb Space Telescope (JWST) have revealed the presence of bright and well-formed galaxies at high redshifts, challenging the predictions of the standard Lambda-Cold Dark Matter (LCDM) cosmological model. This paper explores the potential of Modified Gravity (MOG), specifically Scalar-Tensor-Vector Gravity (STVG), to account for the rapid formation of these galaxies in the early universe. By enhancing the gravitational constant through a dimensionless parameter $\alpha$ and incorporating a massive vector field $\phi_\mu$, MOG predicts deeper gravitational wells that can accelerate the collapse of baryonic matter. We present theoretical insights demonstrating how MOG can facilitate the increase in star formation rate and early formation of galaxies, offering a compelling alternative to LCDM. Our findings suggest that MOG provides a viable framework for understanding the rapid growth of galaxies observed by JWST.\end{abstract}

\section{Introduction}

The discovery of bright and well-formed galaxies at large redshifts only 400 - 500 millions years after the big bang by the James Webb Space Telescope (JWST) has posed significant challenges to our understanding of galaxy formation in the early universe~\cite{Xiao,Labbe}. According to the standard LCDM model, galaxy formation is a gradual process, with small structures merging over time to form larger galaxies. However, the presence of mature galaxies only a few hundred million years after the Big Bang suggests that our current models may be incomplete or require modification.

\begin{figure}
    \centering
    \includegraphics[width=1.0\linewidth]{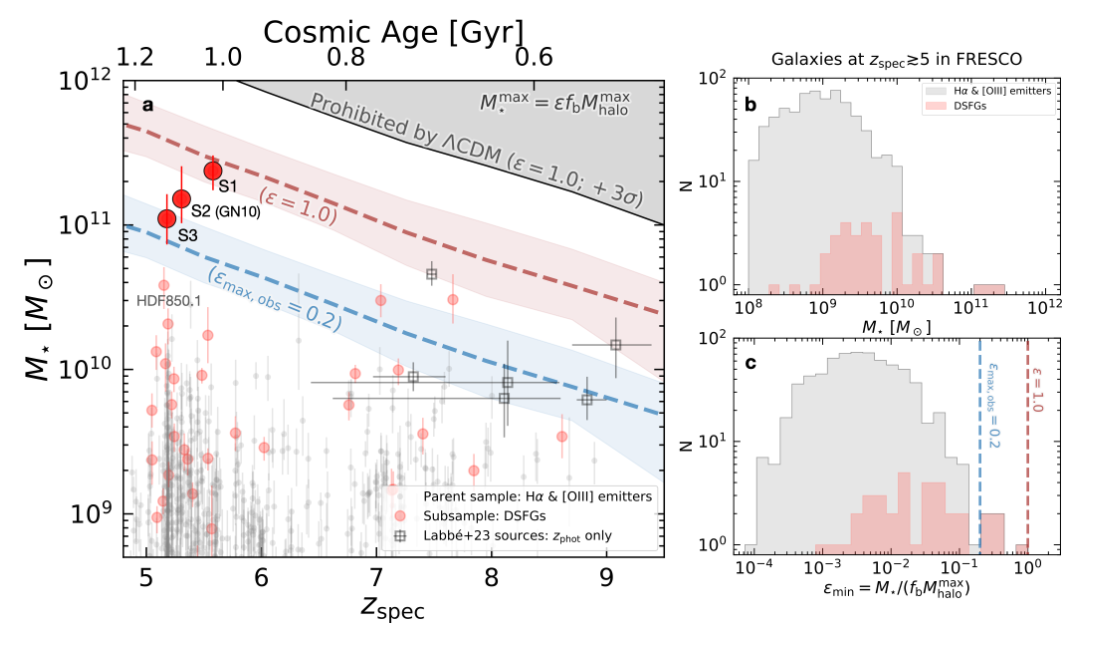}
    \caption{Three outliers, labels $S_1$, $S_2$ and $S_3$, are more massive and have more stars within them than current models can explain. credit: M. Xioa et al., Nature, 635, 311 (2024)}
    \label{fig:enter-label}
\end{figure}

Based on how large and massive the dark matter galaxy halos are expected to be from simulations, stars can only been formed from normal matter with some efficiency. We expect that models that take observations into account should only exhibit of order 20$\%$ efficiency, at maximum. The three galaxies $S_1$, $S_2$, and $S_3$ depicted in Fig. 1, represent a class of objects named 'red monsters'. These galaxies were only about 1 billion years old when they emitted their light. They are very massive with at least 100 billion solar mass stars inside them. They are extremely red in color and rich in dust.

The galaxy $S_1$ forms of order 800 solar masses of new stars per year compared to the Milky Way, which forms about 1 solar mass of new stars per year. Its star formation rate is huge with about a 100$\%$ efficiency rate. These red monster galaxies exhibit no signs of having active galaxy nuclear (AGN) at their centers nor supermassive black holes.

The growth of supermassive black holes (SMBHs) hosted by galaxies and quasars in the early universe is a topic of significant interest and debate. The growth of supermassive black holes (SMBHs) was explored in~\cite{MoffatHaghighi2017}. 
It was demonstrated how MOG offers a solution to the early accretion growth of the SMBHs. An investigation of the dynamics of gas and matter accreting onto the strong gravitational fields of black holes revealed how the strengthening of gravity in MOG rapidly increased the accretion of gas and matter, accelerating the growth of the SMBHs.

The Scalar-Tensor-Vector Gravity STVG or MOG theory, offers a promising alternative to address the problem of early universe galaxy growth. MOG introduces an enhanced gravitational constant, $G = G_N(1+\alpha)$, where $\alpha > 0$ is a dimensionless parameter that strengthens gravity and $G_N$ is Newton's gravitational constant. Additionally, MOG incorporates a massive gravitational vector field $\phi_\mu$ that provides a repulsive force, modifying the dynamics of gravitational interactions at galaxy and cosmological scales.

In this paper, we investigate how MOG can lead to the rapid formation of galaxies in the early universe. We explore the theoretical framework of MOG, focusing on its modifications to gravitational acceleration and potential wells. We demonstrate how these modifications can increase the growth and size of initial matter fluctuations and accelerate the collapse of baryonic matter, leading to the early formation of galaxies. Our results suggest that MOG not only aligns with JWST observations but also provides new insights into the processes driving structure growth, galaxy and star formation.

\section{The MOG weak field approximation}

MOG, also known by the acronym STVG for Scalar-Tensor-Vector Gravity, is a theory of gravitation that involves, in addition to the metric field of general relativity, a gravitational vector field $\phi_\mu$ and scalar degrees of freedom. The core phenomenology of MOG is that while the tensor field couples to matter more strongly than in general relativity, this coupling is partially canceled by a vector field and the resulting repulsion. The vector field is not massless, therefore its range is finite; gravity, therefore, regains its full strength outside the range of the vector force, on galactic and extragalactic scales.

The MOG theory is usually introduced in the form of an action principle and the resulting field equations~\cite{Moffat2006,Moffat2020}. Here we present a brief overview of the non-relativistic, weak field acceleration law.

The gravitational coupling strength $G$ is defined by $G=G_N(1+\alpha)$, and $\alpha$ is a scalar degree of freedom that measures the deviation from General Relativity~\cite{Moffat2006,Moffat2020}. 

 A key premise of MOG is that all baryonic matter possesses, in proportion to its mass $M$, positive gravitational charge $Q_g=\kappa M$, where $\kappa = \sqrt{\alpha G_N}$. This gravitational charge serves as the source of the vector field $\phi_\mu$. MOG fits to galaxy rotation curves, galaxy clusters and cosmology without dark matter have been published~\cite{BrownsteinMoffat2007,MoffatRahvar2013,MoffatRahvar2014,GreenMoffat2019,DavariRahvar2020,IsraelMoffat2018,MoffatHaghighi2017,MoffatTothCMB}.

The modified Newtonian acceleration law for weak gravitational fields and for a point particle can be written
as~\cite{Moffat2006}:
\be
\label{MOGacceleration}
a_{\rm MOG}(r)=-\frac{G_NM}{r^2}[1+\alpha-\alpha\exp(-\mu r)(1+\mu r)].
\ee
This reduces to Newton's gravitational acceleration in the limit
$\mu r\ll 1$. In the limit that $r\rightarrow\infty$, we get,
for approximately constant $\alpha$ and $\mu$:
\be
\label{AsymptoticMOG}
a_{\rm MOG}(r)\approx -\frac{G_N(1+\alpha)M}{r^2}.
\ee

For a distributed baryonic matter source, the MOG weak field acceleration law becomes~\cite{GreenMoffat2019}:
\be 
\textbf{a}_{\rm MOG}=-G_N\int d^3x^\prime\frac{\rho_b(x^\prime)(\textbf{x}-\textbf{x}^\prime)}{|\textbf{x}-\textbf{x}^\prime|^3}
[1+\alpha-\alpha\exp(-\mu|\textbf{x}-\textbf{x}^\prime|)(1+\mu|\textbf{x}-\textbf{x}^\prime|)],
\ee
where $\rho_b$ is the total baryonic mass density. 

\section{Formation of deep potential wells in MOG and matter fluctuations} 

In the context of hierarchical galaxy formation, as dark matter collapses into halos, it creates initial potential wells~\cite{Mo,Henriques,Rodrigues,Angles,
Behroozi}. These potential wells attract baryons (normal matter). As more mass accumulates, the potential well deepens, described by the above equation. Deeper potential wells can trap more baryons, leading to increased galaxy growth.

In cosmological simulations, the Poisson equation, determining the distribution of matter, is solved in comoving coordinates and includes the effects of cosmic expansion:
\be
\label{Poisson}
\nabla^2\phi=4\pi G_N a^2(t)[\rho(x,t) - \bar{\rho}(t)],
\ee
where $\phi$(t) is the gravitational potential, $a(t)$ is the cosmic scale factor and $\bar\rho(t)$ is the mean cosmic density at time $t$.

Equation (\ref{Poisson}) shows how local over-densities in regions where $\rho > \bar\rho$ create deeper potential wells, which in turn attract more matter, leading to the hierarchical growth of structures in the universe. The solution to the equation, combined with equations of hydrodynamics for baryons and the collision-less Boltzmann equation for dark matter, forms the basis for understanding structure formation in the LCDM model.

In modified gravity MOG, the formation of deep gravitational potential wells is a key mechanism that facilitates the rapid collapse of baryonic matter, leading to the accelerated formation of galaxies in the early universe. This section explores the theoretical underpinnings of how MOG modifies gravitational dynamics to create these deep wells.

The MOG gravitational potential $\phi_{\rm MOG}(r)$ is given by
\be
\phi_{\rm MOG}(r)= \frac{G_NM}{r}
\left[1 + \alpha - \alpha\exp(-\mu r)\right]
= \phi_{\rm MOGN} + \phi_Y,
\ee
where 
\be
\phi_{\rm MOGN} = \frac{G_N(1 + \alpha)M}{r},
\ee 
and
\be
\phi_Y = - \alpha G_NM\frac{\exp(-\mu r)}{r}.
\ee
Here, $\mu = 1/r_0$ and $r_0$ is a critical MOG length scale radius of order the sizes of galaxy scales. The escape velocity for the MOG gravitational potential is
\be
v_{\rm esc}(r) = \sqrt{- 2\phi_{\rm MOG}(r)}.
\ee
The MOG Poisson-Helmholtz equation associated with the matter densities is given by~\cite{MoffatToth2013}:
\be
\nabla^2\phi_{\rm MOG} = 4\pi G_N(1 + \alpha)\rho(\textbf {r}) - \mu^2\phi_Y(\textbf {r})
=  4\pi G_N(1 + \alpha)\rho(\textbf {r}) +\alpha\mu^2 G_N\int\frac{\exp(-\mu|\textbf {r} - \textbf {r}^\prime|)}
{|\textbf {r} - \textbf {r}^\prime|}.
\ee

The enhancement of the strength of gravity for $\alpha > 0$ and $\phi_{\rm MOGN} > \phi_Y$ results in a deeper gravitational potential well for a given mass distribution. In the context of cosmological perturbation theory, the growth of density fluctuations 
$\delta(r,t)$ is typically described by a linear growth factor, which evolves according to the gravitational dynamics of the universe~\cite{Schramm,Peebles}. In MOG, the enhanced gravitational constant $G = G_N(1+\alpha)$ modifies the growth rate of these fluctuations.

The equation governing the growth of linear perturbations in MOG can be expressed as~\cite{MoffatToth2013}:
\be
\frac{d^2\delta}{dt^2} - \frac{c_s^2}{a^2}\nabla^2\delta - 4\pi G_N(1 + \alpha)\rho\delta = 0,
\ee
where $\delta$ is the density contrast, defined as 
$\delta=\delta\rho/\rho$, $\rho$ is the baryonic matter density of the universe and $cs$ is the speed of sound. For every Fourier mode, $\delta = 
\delta_k(t)\exp(ik\cdot q)$ and $\nabla^2\delta = -k^2\delta$, we obtain
\be
\label{fluctuations}
\frac{d^2\delta_k}{dt^2} + 2H(t)\frac{d\delta_k}{dt} + \left(\frac{c_s^2k^2}{a^2(t)} 
- 4\pi G_N(1 + \alpha)\rho\right)\delta_k = 0,
\ee
where $H(t)$ is the Hubble parameter and $k/a$ is the co-moving wave number. Solutions of Eq. (\ref{fluctuations}) for large $k$ are dominated by an oscillatory term, while for small $k$ the gravitational growth term dominates.

The presence of $\alpha > 0$ increases 
$G$, leading to a stronger gravitational attraction. This results in a faster growth rate of density fluctuations compared to the standard $\Lambda$CDM model. The equation highlights how MOG modifies the dynamics of initial density fluctuations, potentially explaining the rapid formation of cosmic structures in the early universe. 

In the LCDM hierarchical galaxy formation model, the growth of galaxy mass as the universe expands is a complex process that depends on various factors, including the strength of gravity, cosmic expansion, and the interplay between dark matter and baryons. A model that captures all aspects of this growth, demonstrating the key dependencies is given by the differential equation:
\be
\label{galaxygrowth}
\frac{dM_G}{dt} = \psi (t)\frac{G_N(1 + \alpha)M_G^2}{R_G^2c} 
- \chi (t)H(t)M_G,
\ee
where $M_G$ is the galaxy mass, $R_G$ is the characteristic radius of the galaxy, H(t) is the Hubble parameter and $\psi$ and $\chi$ are time-dependent efficiency factors.

The first term on the right of Eq. (\ref{galaxygrowth}), represents the mass growth due to gravitational attraction. It is proportional to the MOG strength of gravity $G_N(1+\alpha)$ and $M_G^2$ as more massive galaxies attract more matter. The second term on the right, represents mass loss due to cosmic expansion. It is proportional to the Hubble parameter H(t), which describes the expansion rate of the universe.
The $\psi$(t) and $\chi$(t) denote efficiency factors that can account for various astrophysical processes e.g., star formation efficiency, feedback mechanisms and can depend on redshift or cosmic time.

To solve Eq. (\ref{galaxygrowth}) and obtain $M_G(t)$, one would need to specify initial conditions and the forms of $\psi (t)$ and $\chi (t)$, which typically come from more detailed simulations or semi-analytic models. This simplified model demonstrates how the growth of galaxy mass depends on the competition between gravitational attraction proportional to G and cosmic expansion represented by $H(t)$, which are key components 

The enhanced MOG growth rate can lead to the earlier formation structures, such as galaxies and galaxy clusters, which aligns with observations of mature galaxies at high redshifts by the JWST.
 
 \section{Free-fall of molecular gas and galaxy and star formation rate in MOG}
 
 The star formation rate in galaxies is closely related to the free-fall time of gas onto galaxies.

The MOG free-fall time $t_{\rm ff}$ is the characteristic timescale for gravitational collapse of a cloud of gas. It represents how long it would take for a cloud to collapse under its own gravity, if there were no opposing forces. The equation for the MOG free-fall time is:
\be
\label{freefall}
t_{\rm ff} = \left(\frac{3\pi}{32G_N(1+\alpha)
\rho_{\rm gas}}\right)^{1/2},
\ee
where $t_{\rm ff}$ is the free-fall time and $\rho_{\rm gas}$ is the average density of the gas cloud

The star formation rate $S_{\rm FR}$ is generally inversely proportional to the free-fall time. This means that as the free-fall time decreases, the density of gas mass available in a galaxy for star formation rate increases. From Eq. (\ref{freefall}), we can see that the free-fall time is inversely proportional to the square root of the gas density $\rho_{\rm gas}$ and  $G_N(1 + \alpha)$. Higher gas density regions and enhanced gravitational strength have shorter free-fall times, which leads to more rapid star formation.

The actual star formation rate is typically lower than what would be predicted by the free-fall time alone. This is often expressed using an efficiency factor $\epsilon$:
\be
S_{\rm FR}\propto \frac{\epsilon M_{\rm gas}}{t_{\rm ff}},
\ee
where $M_{\rm gas}$ is the mass of available gas.
The efficiency factor $\epsilon$ accounts for various processes that can inhibit star formation, such as turbulence, magnetic fields, and feedback from existing stars.

The Kennicutt-Schmidt law~\cite{Schmidt,Kennicutt} is an empirical relationship that relates the star formation rate surface density to the gas surface density, and can be connected to the free-fall time concept:
\be
\Sigma_{\rm SFR}\propto \Sigma_{\rm gas}^{1/4}.
\ee
This power-law relationship can be understood in terms of the free-fall time, as higher gas densities lead to shorter free-fall times and thus higher star formation rates.

In practice, star formation is more closely correlated with atomic and molecular $H_2$ gas rather than total gas content. The free-fall time of atomic and molecular clouds is particularly relevant for star formation rates. Large-scale galactic processes, such as spiral arm compression or galaxy mergers, can affect local gas densities and thus influence free-fall times and star formation rates across a galaxy.

The free-fall time provides a fundamental timescale for galaxy formation, with shorter free-fall times generally corresponding to higher star formation rates. However, the actual star formation process in galaxies is complex, involving many additional factors beyond simple gravitational collapse. In MOG the galaxy and star formation rate is increased as the free-fall time $t_{\rm ff}$ of the molecular clouds of gas decreases for increasing values of $\alpha$.

The free-fall of atomic and molecular gas onto galaxies, including those around 20 kpc or less in size, contributes to the formation and growth of galaxies and can increase the rate of star formation. Molecular gas is a crucial component for galaxy growth. When it falls onto a galaxy, it increases the galaxy's mass and provides the raw material for various processes within the galaxy.

Molecular gas, particularly cold molecular $H_2$ hydrogen, is the primary fuel for star formation. As this gas falls into the galaxy, it can increase the overall gas density in certain regions, which is a key factor in star formation. It triggers gravitational instabilities, leading to the collapse of gas clouds and subsequent star formation. Interaction with existing gas in the galaxy can potentially create shock waves that compress gas and initiate star formation.

For galaxies around $r_0\sim 20$ kpc or less in size, the impact of in-falling molecular gas can be significant. It can replenish the gas supply in the outer regions of the galaxy. It may contribute to the formation of structures like extended disks or rings. In some cases, it could trigger star formation in the outer regions of the galaxy.

In some scenarios, particularly in galaxy clusters, the in-fall of cooling gas from the intergalactic medium can lead to what is known as a "cooling flow." This can provide a steady supply of molecular gas to the galaxy.

It is important to note that the in-fall process is regulated by various feedback mechanisms, such as:
supernova feedback, active galactic nucleus (AGN) feedback and galactic winds. These processes can heat up or expel gas, potentially reducing the efficiency of gas infall and star formation. The exact impact depends on various factors, including the galaxy's environment, its current gas content, and its dynamical state. The process is complex and interconnected with many other aspects of galaxy evolution.

\section{Conclusions}

The observations of bright, massive and well-formed galaxies at high redshifts by the JWST have posed significant challenges to the standard LCDM cosmological model, which predicts a more gradual formation of galaxies. In this paper, we have explored the potential of modified gravity (MOG), specifically Scalar-Tensor-Vector Gravity (STVG), to account for the rapid formation of these galaxies in the early universe.

Our analysis demonstrates that MOG, through the enhancement of the gravitational strength and the inclusion of a massive vector field, can create deeper gravitational potential wells. These wells facilitate the accelerated collapse of baryonic matter, leading to the early formation of galaxies. The modified gravitational dynamics in MOG provide a compelling framework for understanding the rapid growth of cosmic structures observed by JWST.

We have presented theoretical insights and equations describing the enhanced gravitational acceleration and the growth of density fluctuations in MOG. Our findings suggest that the increased gravitational strength due to $\alpha > 0$ leads to a faster growth rate of initial density fluctuations, potentially explaining the presence of mature galaxies at high redshifts. We have also investigated the rate of star formation in MOG. The faster star formation results from the faster time of free-fall $t_{\rm ff}$ of inter-galactic medium gas onto galaxies.

In conclusion, while MOG offers a promising alternative to LCDM, further research is needed to fully understand its implications for early universe galaxy growth and formation. Future work should include detailed numerical simulations to explore the parameter space of MOG, as well as comparisons with future observational data from JWST and other telescopes.

\section*{Acknowledgments}

I thank Martin Green and Viktor Toth for helpful discussions. Research at the Perimeter Institute for Theoretical Physics is supported by the Government of Canada through industry Canada and by the Province of Ontario through the Ministry of Research and Innovation (MRI).

\end{document}